\documentstyle[12pt]{article}
\begin{document}

\thispagestyle{empty}
\begin{flushright}
{ \bf DFTUZ/94/05}
\\{\large \bf hep-th/9404081}
\end{flushright}
\vskip 2truecm

\begin{center}

{ \large \bf  SUPERSYMMETRY WITHOUT FERMIONS}\\
\vskip0.8cm
{ \bf
Mikhail S. Plyushchay\footnote{E-mail: mikhail@cc.unizar.es}}\\[0.3cm]
{\it Departamento de Fisica Teorica, Facultad de Ciencias}\\
{\it Universidad de Zaragoza, 50009 Zaragoza, Spain}\\
{\it and}\\
{\it Institute for High Energy Physics, Protvino, Russia}\\[0.5cm]
\begin{center}
April 1994
\end{center}

\vskip1.5cm

                            {\bf Abstract}
\end{center}
The simplest $N=2$ supersymmetric quantum mechanical system
is realized in terms of the bosonic creation and annihilation
operators obeying either ordinary or deformed Heisenberg algebra
involving Klein operator. The construction comprises both the exact
and spontaneously broken supersymmetry cases with the scale of
supersymmetry breaking being governed by the deformation parameter.
Proceeding from the broken supersymmetry case, we realize the Bose-Fermi
transformation and obtain spin-$1/2$ representation of $SU(2)$ group in
terms of one bosonic oscillator. We demonstrate that the constructions
can be generalized to the more complicated $N=2$ supersymmetric systems,
in particular, corresponding to the Witten supersymmetric quantum mechanics.

\newpage

It was demonstrated recently \cite{ply} that the classical analog of the
quantum massless superparticle can be constructed without incorporating odd
Grassmann variables
being the classical analogs of the corresponding fermionic operators.
Such ``bosonization"
was realized with the help of topologically nontrivial classical variables,
which after quantization supply the system with the necessary fermionic
degrees of freedom.
But, moreover, the next step can be carried out
in the direction of bosonization of the supersymmetry.
In recent paper \cite{mac}, devoted to the generalized harmonic
oscillator systems, Brzezinski, Egusquiza and Macfarlane
have pointed out on a remarkable
possibility of uncovering supersymmetry structure totally described
within the Fock space of the bosonic oscillator.
In the present letter we shall investigate such a possibility of
realizing $N=2$ supersymmetry for the case of quantum mechanical bosonic
oscillator revealing
similarity of the constructions with the standard ones comprising fermionic
degrees of freedom.
Certainly, the bosonization mechanisms for the supersymmetry could find
their
concrete interesting  applications similarly to the Bose-Fermi transmutaion
constructions,
known since very long time ago \cite{jor}, which nowdays together with the
anyonic generalizations are applied, e.g.,
to the description of the planar physical  phenomena,
viz. high-$T_{c}$ superconductivity and fractional quantum Hall
effect \cite{wil}.

We shall start with the case of the ordinary bosonic oscillator and
realize $N=2$ supesymmetry in terms of its creation and annihilation
operators demonstrating that there are two possibilities corresponding
to the cases of exact and spontaneuosly broken supersymmetry.
Proceeding from
the broken supersymmetry case, we shall construct the fermionic oscillator
variables, i.e. realize a Bose-Fermi transformation in terms of only one
bosonic oscillator. The fermionic oscillator variables will permit us
to represent the Hamiltonian for both cases in the form similar to the
standard supersymmetric form and
construct spin-1/2 representation of the $SU(2)$ group
on the one bosonic oscillator Fock space.
Then we shall generalize all the constructions to the case of the
bosonic oscillator operators obeying the deformed Heisenberg algebra
involving the Klein operator \cite{vas,mac}. The interesting feature of
such generalization
is that in the case of the spontaneously broken supersymmetry
the scale of supersymmetry breaking is defined by the deformation parameter
of the  system. After that we shall show how the described constructions
can be generalized
to the more complicated $N=2$ supersymmetric systems, in particular,
corresponding to the Witten supersymmetric quantum mechanics.

So, let us consider
the Fock space of states for the ordinary bosonic oscillator formed by the
complete set of orthonormal vectors
\begin{equation}
|n>=C_{n}(a^{+})^{n}|0>, \quad
<n|n'>=\delta_{nn'},\quad n=0,1,\ldots,
\label{n0}
\end{equation}
constructed over the  vacuum state defined by the relations:
\begin{equation}
a^{-}|0>=0,\quad <0|0>=1,
\label{vac}
\end{equation}
where $a^{+}$ and $a^{-}$ are creation and annihilation operators
satisfying the algebra
\begin{equation}
[a^{-},a^{+}]=1,
\label{heis}
\end{equation}
and the normalization constants can be chosen as
$C_{n}=(n!)^{-1/2}$.
Introduce the Klein operator $K$ defined by the relations
\begin{equation}
K^{2}=1,\quad \{K,a^{\pm}\}=0,
\label{kle}
\end{equation}
which separates the complete set of states $|n>$ into even and odd
subspaces:
\begin{equation}
K|n>=\kappa\cdot (-1)^{n}|n>.
\label{z2}
\end{equation}
The sign factor $\kappa$ can be removed here by redefining the vacuum
state and without loss of generality we can put
$\kappa=+1$.
The operator $K$ can be realized in the form
$
K=\exp{i\pi N},
$
or, in the explicitely hermitian form
\begin{equation}
K=\cos \pi N,
\label{kcos}
\end{equation}
with the help of the number operator, $N=a^{+}a^{-}$,
\begin{equation}
N|n>=n|n>.
\label{num}
\end{equation}

Consider now  an operator $Q^{+}$
together with hermitian conjugate operator $Q^{-}=(Q^{+})^{\dagger}$ of
the most general form
linear in the oscillator variables $a^{\pm}$ with coefficients depending
on the Klein operator $K$,
\[
Q^{+}=\frac{1}{2}a^{+}(\alpha+\beta K)+\frac{1}{2}a^{-}(\gamma+\delta K),
\]
where $\alpha$, $\beta$, $\gamma$ and $\delta$ are arbitrary complex
numbers,
and demand that they would be nilpotent operators:
$
Q^{+2}=Q^{-2}=0.
$
This requirement leads to the restriction:
$
\beta=\epsilon\alpha,\quad \delta=\epsilon\gamma,
$
where $\epsilon$ is a sign parameter.
Therefore, there are two possibilities for choosing  operator $Q^{+}$:
\[
Q^{+}_{\epsilon}=(\alpha a^{+}+\gamma a^{-})\Pi_{\epsilon}, \quad
\epsilon=\pm,
\]
where by $\Pi_{\epsilon}$ we denote the hermitian operators
\[
\Pi_{\pm}=\frac{1}{2}(1\pm K)
\]
being the projectors:
$\Pi_{\pm}^{2}=\Pi_{\pm},$ $\Pi_{+}\Pi_{-}=0,$ $\Pi_{+}+\Pi_{-}=1.$
The anticommutator $\{Q^{+}_{\epsilon},Q^{-}_{\epsilon}\}$
has here the form:
\[
\{Q^{+}_{\epsilon},Q^{-}_{\epsilon}\}=
a^{+2}\alpha\gamma^{*}+a^{-2}\alpha^{*}\gamma
+\frac{1}{2}\left(\{a^{+},a^{-}\}-\epsilon K
[a^{-},a^{+}]\right)(\alpha\alpha^{*}+\gamma\gamma^{*}).
\]
Whence we conclude that the anticommutator will commute with the number
operator $N$ if we choose parameters in such a way that
$\alpha\gamma^{*}=0$.
Putting $\alpha=0$, normalize the second parameter as $\gamma=e^{i\varphi}$.
But since the phase factor can be removed by the unitary transformation
of the oscillator operators $a^{\pm}$, we arrive at the nilpotent
operators in the very compact form similar to that of the
supercharges for the simplest $N=2$
supersymmetric systems in the standard approach including fermionic
operators
(see below eq. (\ref{qpm})):
\begin{equation}
Q^{+}_{\epsilon}=a^{-}\Pi_{\epsilon},\quad
Q^{-}_{\epsilon}=a^{+}\Pi_{-\epsilon}.
\label{qexp}
\end{equation}
They together with the operator
\begin{eqnarray}
H_{\epsilon}&=&\frac{1}{2}\{a^{+},a^{-}\} -\frac{1}{2}\epsilon K [a^{-},
a^{+}]
\label{hexp1}\\
&=&\frac{1}{2}\{a^{+},a^{-}\} -\frac{1}{2}\epsilon K
\label{hexp2}
\end{eqnarray}
form the $N=2$ superalgebra:
\begin{equation}
Q^{\pm2}_{\epsilon}=0,\quad \{Q^{+}_{\epsilon},Q^{-}_{\epsilon}\}
=H_{\epsilon},\quad
[Q^{\pm}_{\epsilon},H_{\epsilon}]=0.
\label{salg}
\end{equation}
Let us notice that the hermitian supercharge operators
$Q^{1,2}_{\epsilon}$,
\begin{equation}
Q^{\pm}_{\epsilon}=\frac{1}{2}(Q^{1}_{\epsilon}\pm iQ^{2}_{\epsilon}),
\quad
\{Q^{i}_{\epsilon},Q^{j}_{\epsilon}\}=2\delta^{ij}H_{\epsilon},
\label{qqq}
\end{equation}
have the following compact form:
\begin{equation}
Q^{1}_{\epsilon}=\frac{1}{\sqrt{2}}(x+i\epsilon pK),\quad
Q^{2}_{\epsilon}=\frac{1}{\sqrt{2}}(p-i\epsilon xK)=-i
\epsilon Q^{1}_{\epsilon}K
\label{q12}
\end{equation}
in terms of the  bosonic operators of coordinate $x$ and momentum
$p$, $a^{\pm}=(x\mp ip)/\sqrt{2},$  $[x,p]=i.$

Consider the spectrum of the supersymmetric Hamiltonian
(\ref{hexp2}). In the case when $\epsilon=-$,
the states $|n>$ are the eigenstates  of the operator $H_{-}$ with
the eigenvalues
\begin{equation}
E_{n}^{-}=2[n/2]+1,
\label{eb}
\end{equation}
where $[n/2]$ means the integer part of $n/2$. Therefore, for $\epsilon=-$
we have the case of the spontaneously  broken supersymmetry with
$E_{n}^{-}>0$
and all the states
$|n>$ and $|n+1>$, $n=2k$, $k=0,1,\ldots$, are paired in supermultiplets.
For $\epsilon=+$ we have the case of the exact supersymmetry characterized
by the spectrum
\begin{equation}
E_{n}^{+}=2[(n+1)/2],
\label{ee}
\end{equation}
i.e. here the vacuum state with $E_{0}^{+}=0$ is a supersymmetry singlet,
whereas $E_{n}^{+}=E_{n+1}^{+}>0$ for  $n=2k+1, k=0,1,\ldots$.
Thus, we have demonstrated that one can realize both cases of the
spontaneously broken
and exact $N=2$ supersymmetries with the help of only one bosonic
oscillator.

Due to the property $E_{n}^{-}>0$ taking place for $\epsilon=-$,
we can construct the Fermi oscillator operators:
\begin{eqnarray}
f^{\pm}&=&\frac{Q^{\mp}_{-}}{\sqrt{H_{-}}}
\label{fdef}\\
&=&a^{\pm}\cdot \frac{\Pi_{\pm}}{\sqrt{N+\Pi_{+}}},
\nonumber
\end{eqnarray}
\begin{equation}
\{f^{+},f^{-}\}=1,\quad f^{\pm 2}=0,
\label{ff}
\end{equation}
i.e. one can realize the Bose-Fermi transformation in terms of one
bosonic oscillator.
Notice here, that though, obviously,
operators $a^{\pm}$ do not commute with $f^{\pm}$, nevertheless,
the operator $H_{\epsilon}$ can be written
in the form of the simplest supersymmetric Hamiltonian:
\begin{equation}
H_{\epsilon}=\frac{1}{2}\{a^{+},a^{-}\}+\epsilon \frac{1}{2}[f^{+},f^{-}].
\label{hb}
\end{equation}
Indeed, let us compare the constructed systems, being distinguished by the
sign parameter $\epsilon$, with the simplest ordinary $N=2$ supersymmetric
systems described
in terms of mutually commuting bosonic ($a^{\pm}$) and fermionic
($b^{\pm}$) oscillator operators:
\begin{equation}
\{b^{+},b^{-}\}=1,\quad
b^{\pm2}=0,\quad
[a^{\pm},b^{\pm}]=0,
\label{ab}
\end{equation}
with the supercharge and Hamiltonian operators realized as
\begin{equation}
Q^{+}_{\epsilon}=a^{-}b^{\epsilon},\quad Q^{-}_{\epsilon}=
a^{+}b^{-\epsilon},\quad
\epsilon=\pm,
\label{qpm}
\end{equation}
\begin{equation}
H_{\epsilon}=\frac{1}{2}\{a^{+},a^{-}\}+\epsilon \frac{1}{2}[b^{+},b^{-}].
\label{hpm}
\end{equation}
Here  $\epsilon=+$ and $\epsilon=-$ also give
the cases of the exact and spontaneously
broken supersymmetries, and we see that the Hamiltonian (\ref{hb})
formally coincides with $H_{\epsilon}$ for the  system
(\ref{ab})--(\ref{hpm}).
It is just from such reasons of similarity we have chosen the signs
in both sides of eq. (\ref{fdef}).

Having  fermionic oscillator variables $f^{\pm}$,
one can construct the operators:
\begin{equation}
S_{1}=\frac{1}{2}(f^{+}+f^{-}),\quad
S_{2}=-\frac{i}{2}(f^{+}-f^{-}),\quad
S_{3}=f^{+}f^{-}-1/2,
\label{s123}
\end{equation}
satisfying the $su(2)$ algebra:
\begin{equation}
[S_{i},S_{j}]=i\epsilon_{ijk}S_{k},\quad i,j,k=1,2,3.
\label{ss}
\end{equation}
Due to the relation
$S_{i}S_{i}=3/4,$
it means that we can realize unitary spin-$1/2$ representaion of
of the $SU(2)$ group acting on every 2-dimensional space of states
$(|2k>,|2k+1>)$,
$k=0,1,\ldots$, in an irreducible way.

Now let us go to the generalizations of these simple constructions,
and first consider the case of the bosonic oscillator variables satisfying
the deformed Heisenberg algebra \cite{vas,mac}
\begin{equation}
[a^{-},a^{+}]=1+\nu K.
\label{def}
\end{equation}
Here the Klein operator $K$ is also given by the relations of the
form (\ref{kle}).
Let us introduce again the vacuum state defined by eqs.
(\ref{vac}) and put $\kappa=+1$ in the relation of the form (\ref{z2}).
Then we find that the operator $a^{+}a^{-}$ acts on the states $|n>$
defined by  eqs. (\ref{n0})  in the following way:
\begin{equation}
a^{+}a^{-}|n>=\left(n+\frac{\nu}{2}\left(1+(-1)^{n+1}\right)\right)|n>.
\label{apm}
\end{equation}
{}From here we conclude that in the case when
\begin{equation}
\nu >-1,
\label{restr}
\end{equation}
the space of unitary representation of  algebra
(\ref{def}), (\ref{kle}) is given by the complete set of the orthornormal
states
(\ref{n0}) with the coefficients
\[
C_{n}=\left(\Pi_{l=1}^{n}\left(l+\frac{\nu}{2}\left(1+(-1)^{n+1}\right)
\right) \right)^{-1/2}.
\]

{}From eqs. (\ref{apm}) and (\ref{z2}) we get the equality:
$a^{+}a^{-}=N+\nu(1-K)/2,$
where through $N$ we denoted the number operator
(\ref{num}). Then, using the previous relation and eq. (\ref{def}), we get
$a^{-}a^{+}=N+1+\nu(1+K)/2,$
and from these two relations we arrive as a result at the following
 expression
for the number operator in terms of the operators $a^{\pm}$:
\begin{equation}
N=\frac{1}{2}\{a^{-},a^{+}\}-\frac{1}{2}(\nu+1).
\label{ndef}
\end{equation}
Now, we can realize the Klein operator $K$ in terms of the operators
$a^{\pm}$ by means of equalities (\ref{kcos}) and (\ref{ndef}), and
the constructions performed with the use of ordinary bosonic oscillator
operators can be repeated here in the same way. So,
we get the supercharges  and supersymmetry Hamiltonian in the forms
(\ref{qexp}) and
(\ref{hexp1}), respectively.
Then we find that in the case $\epsilon=+$ we have again the
exact supersymmetry and the states $|n>$ are the eigenstates of the
Hamiltonian $H_{+}$ with the same spectrum (\ref{ee}) as in the case of
Heisenberg
algebra (\ref{heis}), whereas in the case $\epsilon=-$,
we have the case of the spontaneously broken supersymmetry
with the shifted energy spectrum:
instead of (\ref{eb}) we have here
\begin{equation}
E_{n}^{-}=2[n/2]+1+\nu.
\label{scale}
\end{equation}
Therefore, in this case the shift of the energy (the scale of the
supersymmetry breaking)
is defined by the deformation
parameter, but $E_{n}^{-}>0$ for all $n$  due to the restriction
(\ref{restr}).
It means that in the case of the deformed bosonic oscillator we can realize
again
the Bose-Fermi transformation and get spin-$1/2$ reperesentation of
$SU(2)$  with the help of the
relations of the same form (\ref{fdef}), (\ref{ff}), (\ref{s123}) and
(\ref{ss})
as in the case of the ordinary oscillator.
Therefore, from the point of view of the supersymmetry constructions
the deformation of the Heisenberg algebra reveals itself in the scale of
supersymmetry breaking.

Let us show how the previous constructions can be generalized to the case
corresponding
to the more complicated quantum mechanical supersymmetric
systems \cite{gen}. To this end, consider the operators
\begin{equation}
\tilde{Q}^{\pm}_{\epsilon}=A^{\mp}\Pi_{\pm\epsilon}
\label{tilq}
\end{equation}
with odd mutually conjugate operators $A^{\pm}=A^{\pm}(a^{+},a^{-})$,
$A^{-}=(A^{+})^{\dagger}$, $KA^{\pm}=-A^{\pm}K$.
These properties of $A^{\pm}$ guarantee that the
operators
$\tilde{Q}^{\pm}_{\epsilon}$ are, in turn, mutually conjugate,
$\tilde{Q}^{-}_{\epsilon}=
(\tilde{Q}^{+}_{\epsilon})^{\dagger}$,
and nilpotent:
\begin{equation}
(\tilde{Q}{}^{\pm}_{\epsilon})^{2}=0.
\label{tqq}
\end{equation}
Taking the anticommutator
\begin{equation}
\tilde{H}_{\epsilon}=\{\tilde{Q}^{+}_{\epsilon},\tilde{Q}^{-}_{\epsilon}\}
\label{the}
\end{equation}
as the Hamiltonian,
we get the  $N=2$ superalgebra defined by the relations
(\ref{tqq}), (\ref{the}) and by the commutator
\begin{equation}
[\tilde{H}_{\epsilon},\tilde{Q}^{\pm}_{\epsilon}]=0.
\label{thq}
\end{equation}
The explicite form of the supersymmetric Hamiltonian here is
\begin{equation}
\tilde{H}_{\epsilon}=\frac{1}{2}\{A^{+},A^{-}\}-\frac{1}{2}\epsilon K
[A^{-},A^{+}].
\label{htilde}
\end{equation}
Choosing the operators $A^{\pm}$, e.g., in the form
\[
A^{\pm}=\frac{1}{\sqrt{2}}(\mp ip+W(x))
\]
with $W(-x)=-W(x)$, in the case of undeformed Heisenberg algebra
(\ref{heis})
we get for supesymmetric Hamiltonian (\ref{htilde}), in its turn,
the form
\[
\tilde{H}_{\epsilon}=\frac{1}{2}(p^{2}+W^{2}-\epsilon K W')
\]
corresponding to the Witten supesymmetric quantum mechanics  \cite{wit}
with odd superpotential $W$.

Thus, we see that $N=2$ supersymmetry is realizable in terms of only one
bosonic oscillator in the form very similar to that of the
standard approach comprising fermionic oscillator operators.
In the case of the simplest supersymmetric system considered above
the Hamiltonian (\ref{hexp2})
in Schr${\rm \ddot{o}}$dinger representation takes the form:
\[
H_{\epsilon}=\frac{1}{2}\left(-\frac{d^{2}}{dx^{2}}+x^{2}\right)-
\frac{1}{2}\epsilon
\sin \frac{\pi}{2}\left(x^{2}-\frac{d^{2}}{dx^{2}}\right).
\]
Therefore, the supersymmetrization of the pure bosonic system is
conditioned here on the
specific nonlocal character of the Hamiltonian.

To conclude, let us point out two problems to be interesting for further
consideration.
The first one concerns possible generalizations to the cases
of supersymmetric $N>2$ and parasupersymmetric  quantum mechanical systems.
For the purpose one could try to use the generalizations
of the Klein operator of the form: $\tilde{K}{}^{n}=1$, $n>2$.
The second, obvious problem consists in generalizing the bosonization
of the $N=2$
supersymmetry to the quantum field systems, $(1+1)$-dimensional in the
simplest case.

$\ $

I am grateful to I.A. Bandos, D.P. Sorokin and D.V. Volkov
for discussions
stimulated the appearance of this work. The research was partially
supported by MEC-DGICYT (Spain).


\newpage

\begin{thebibliography}{**}
\bibitem{ply}
M.S. Plyushchay, {\it Phys. Lett.} {\bf B280} (1992)  232

\bibitem{mac}
T. Brzezinski, I.L. Egusquiza and A.J. Macfarlane,
{\it Phys. Lett.} {\bf B311} (1993) 202

\bibitem{jor}
P. Jordan and E.P. Wigner, {\it Z. Phys.} {\bf 47} (1928) 631

\bibitem{wil}
F. Wilczek, {\it Fractional Statistics and Anyon Superconductivity},
(World Scientific, Singapore, 1990)

\bibitem{vas}
M.A. Vasiliev, {\it Pis'ma JETP} {\bf 50} (1989) 344;
{\it Inter. J. Mod. Phys.} {\bf A6} (1991) 1115;\\
L. Brink, T.H. Hansson T.H. Konstein and M.A. Vasiliev,
{\it Nucl. Phys.} {\bf B401} (1993) 591


\bibitem{gen}
L.E. Gendenshtein and I.V. Krive, {\it Sov. Phys. Usp.} {\bf 28} (1985) 645

\bibitem{wit}
E. Witten, {\it Nucl. Phys.} {\bf B188} (1981) 513


\end{thebibliography}
\end{document}